\documentclass[aps,prb,twocolumn,superscriptaddress]{revtex4-1}
\usepackage{bm}
\usepackage{mathtools}
\usepackage{graphicx}% Include figure files
\usepackage{graphics}% Include figure files
\usepackage{dcolumn}% Align table columns on decimal point
\usepackage{float}
\usepackage{epsfig}
\usepackage{epstopdf} %converting to PDF
\usepackage{color}
\usepackage{amsmath}
\begin{document}
	\title{An electrically driven spin qubit based on valley mixing}
	\author{Wister Huang}
	\email[E-mail: ]{wister.huang@student.unsw.edu.au}
	\affiliation{Center for Quantum Computation and Communication Technology, School of Electrical Engineering and Telecommunications, The University of New South Wales, Sydney, New South Wales 2052, Australia}
	
	\author{Menno Veldhorst}
	\email[E-mail: ]{m.veldhorst@tudelft.nl}
	\affiliation{Center for Quantum Computation and Communication Technology, School of Electrical Engineering and Telecommunications, The University of New South Wales, Sydney, New South Wales 2052, Australia}
	\affiliation{QuTech, TU Delft, 2600 GA Delft, The Netherlands}
	
	\author{Neil M. Zimmerman}
	\affiliation{National Institute of Standards and Technology, Gaithersburg , Maryland, USA, 20899}
	
	\author{Andrew S. Dzurak}
	\email[E-mail: ]{a.dzurak@unsw.edu.au}
	\affiliation{Center for Quantum Computation and Communication Technology, School of Electrical Engineering and Telecommunications, The University of New South Wales, Sydney, New South Wales 2052, Australia}
	
	\author{Dimitrie Culcer}
	\email[E-mail: ]{d.culcer@unsw.edu.au}
	\affiliation{School of Physics, The University of New South Wales, Sydney 2052, Australia}
	\date{\today}
\begin{abstract}
The electrical control of single spin qubits based on semiconductor quantum dots is of great interest for scalable quantum computing since electric fields provide an alternative mechanism for qubit control compared with magnetic fields and can also be easier to produce. Here we outline the mechanism for a drastic enhancement in the electrically-driven spin rotation frequency for silicon quantum dot qubits in the presence of a step at a hetero-interface. The enhancement is due to the strong coupling between the ground and excited states which occurs when the electron wave-function overcomes the potential barrier induced by the interface step. We theoretically calculate single qubit gate times $t_{\pi}$ of 170 ns for a quantum dot confined at a silicon/silicon-dioxide interface. The engineering of such steps could be used to achieve fast electrical rotation and entanglement of spin qubits despite the weak spin-orbit coupling in silicon.
\end{abstract}
		\maketitle

\section{Introduction}

Solid state spin qubits based on quantum dots~\cite{PhysRevA.57.120} take a variety of forms~\cite{Bluhm2011, VeldhorstM.2014, KawakamiE.2014, Veldhorst2015, PhysRevLett.115.106802, Petta2180, PhysRevB.82.075403, Enge1500214, PhysRevB.91.205434, Kim2015}. Silicon is an ideal host for spin qubits thanks to the absence of piezoelectric electron-phonon coupling, to nuclear-spin free isotopes ~\cite{RevModPhys.85.961,MRC:9492360} enabling isotopic purification to remove the hyperfine coupling, and compatibility with industrial manufacturing technologies. Recent experiments have realized high-fidelity single-qubit operations ~\cite{VeldhorstM.2014} and two-qubit logic gates ~\cite{Veldhorst2015} in silicon metal-oxide-semiconductor (Si-MOS) dots in isotopically enriched $^{28}$Si, while high fidelity single-qubit operations have been achieved in Si/SiGe dots in both $^{28}$Si ~\cite{Enge1500214} and naturally occurring Si~\cite{KawakamiE.2014,2016arXiv160207833T}.

Fast, individually addressable qubit operations are essential for scalable architectures. Since electric fields can be easier to produce and control locally than magnetic fields, rotating electron spins electrically could not only be faster, but would also facilitate scalability. A significant effort has therefore focused on achieving electron dipole spin resonance (EDSR) of single spins in quantum dots. Experimentally this relies on spin-orbit coupling, which allows simultaneous changes of both the orbital and spin states, an AC electric field driving purely orbital transitions, and a static, uniform magnetic field needed to break time reversal. Rabi frequencies $f\approx$ 3-4 MHz have been achieved in GaAs ~\cite{Nowack1430,2016arXiv160302829R} In silicon, in which the electron spin-orbit coupling is weak, fast EDSR requires the inhomogeneous magnetic field of a nanomagnet, and $f \approx$ 4 MHz has been realized in Si/SiGe qubits ~\cite{KawakamiE.2014,1882-0786-8-8-084401}. 

In this work we show that spin-orbit induced EDSR in silicon is strongly enhanced by the combination of two ubiquitous features of silicon quantum dots: the valley degree of freedom and steps at the silicon interface, which can be either identified or engineered. An interface step leads to strong coupling between ground and excited orbital and valley states and, through the spin-valley coupling provided by the spin-orbit interaction, a large enhancement of EDSR can occur when the electron wavefunction is positioned in a small region near the step. This implies that spin-orbit coupling can be used as an intrinsic mechanism for EDSR in silicon, and its impact should also be considered in nanomagnet-based spin qubits. We consider in detail dots formed at Si/SiO$_2$ interfaces, but we note that the mechanism applies also to Si/SiGe quantum dot qubits.

The conduction band minima in Si/SiO$_2$ heterostructures grown along (001) lie in two equivalent valleys perpendicular to the interface at $ \pm k_0 = \pm 0.85 \, (2\pi/a_{Si})$, \cite{RevModPhys.85.961} with the Si lattice constant $a_{Si} \approx 5.43 \AA$. The sharp interface potential and $\hat{\bm z}$-direction ($\parallel$[001]) electric field give rise to a valley-orbit coupling~\cite{PhysRevB.82.205315}, whose magnitude is responsible for the several hundred $\mu eV$ splittings between valley eigenstates observed experimentally.\cite{Yang2013} Spin-orbit coupling in Si has both intravalley and intervalley terms, \cite{PhysRevB.92.201401,PhysRevB.73.235334} and tuning the valley-orbit coupling has a noticeable effect on spin dynamics. Experimental studies have shown the effective $g$-factor is modified by an out-of-plane electric field in both valley eigenstates in silicon, confirming the theoretical predictions of $g$-factor sensitivity to valley composition. \cite{PhysRevB.92.201401}

\begin{figure}
	\includegraphics[width= \linewidth]{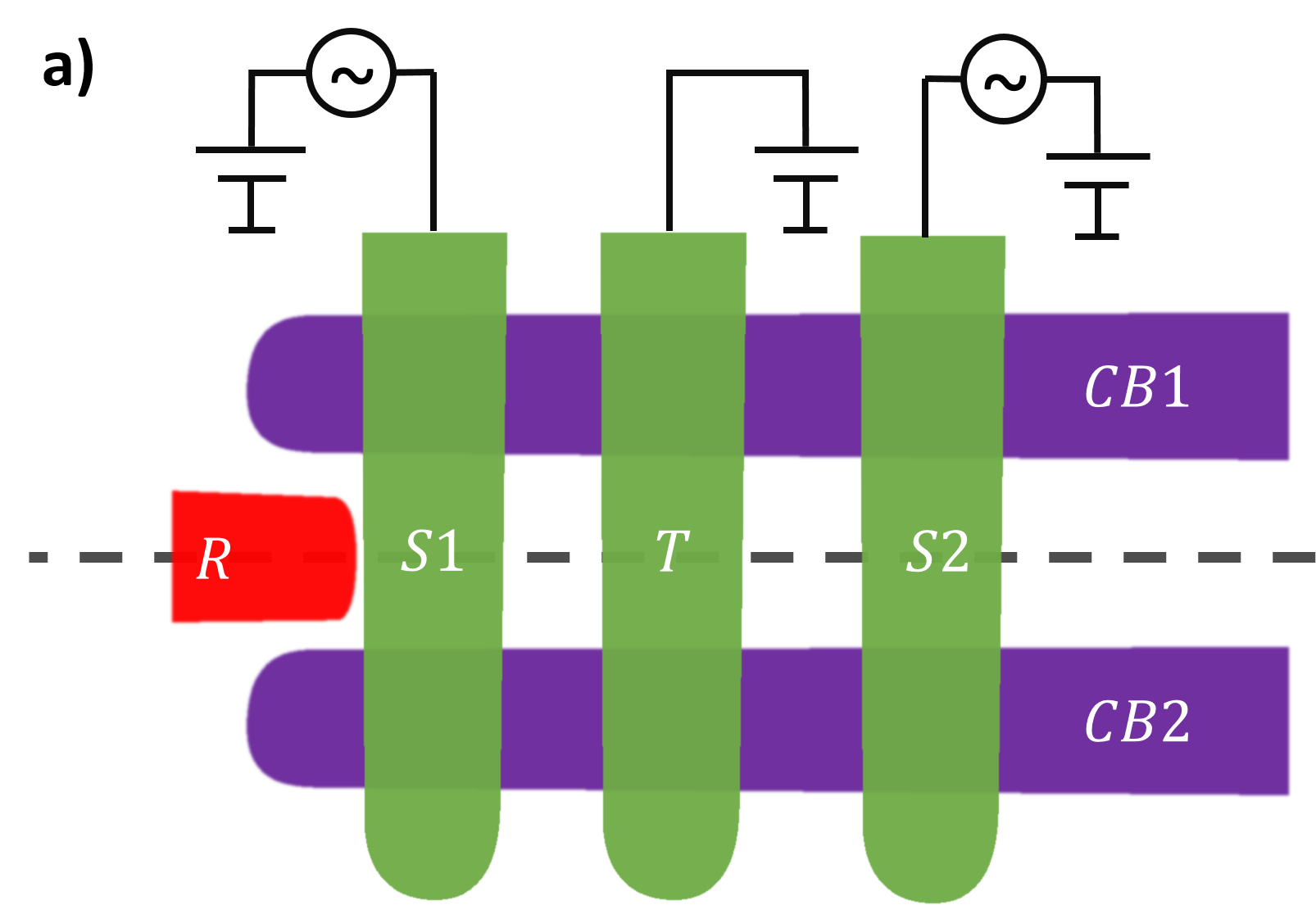}
	\includegraphics[width= \linewidth]{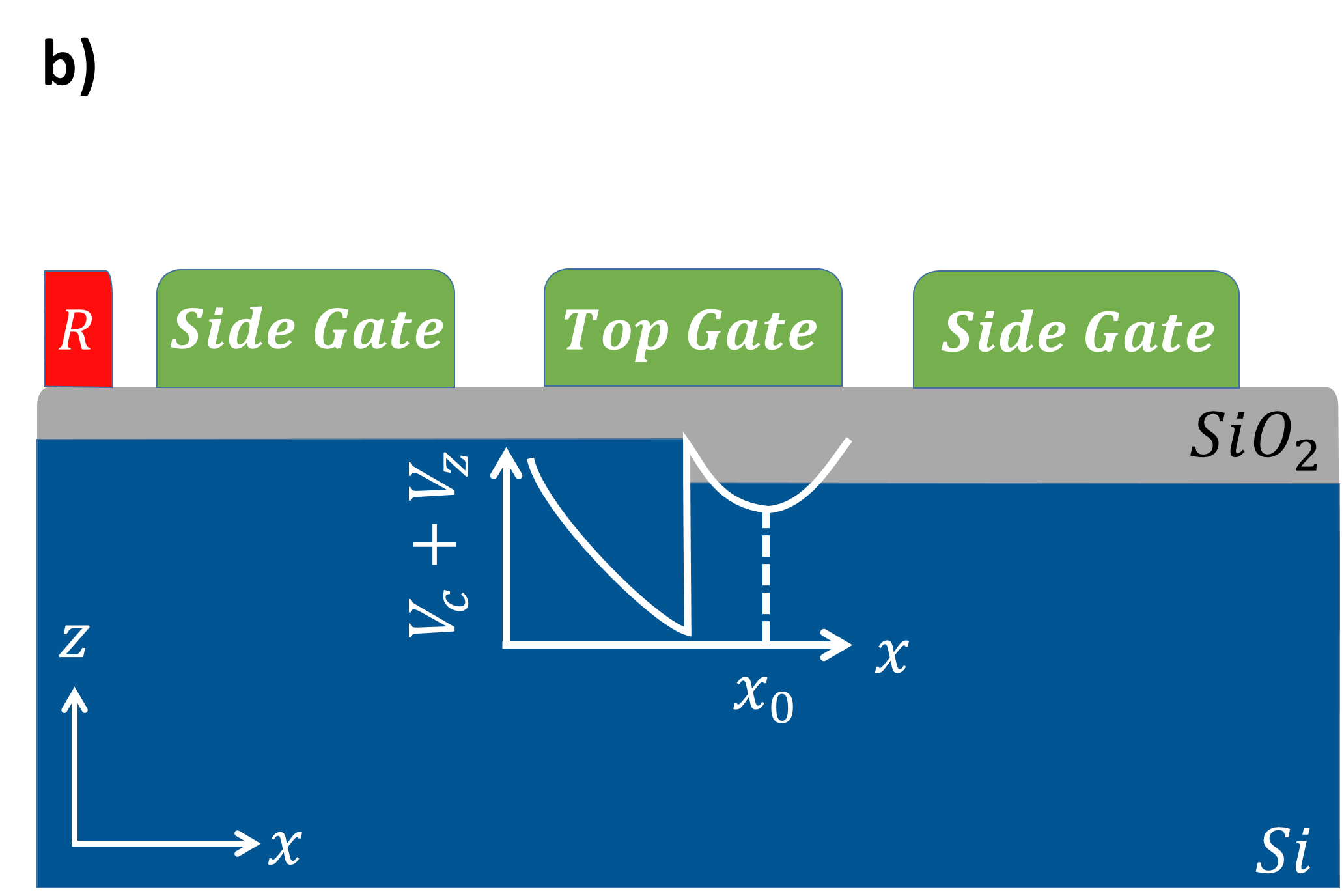}
	\caption{(a) Top view and (b) cross sectional schematic representation of a multi-gated metal oxide semiconductor structure with a single interface step of height $d$. The dot is defined by the confinement barriers (CB) and located beneath the plunger gates. Two side gates can produce both DC and AC in plane electric fields to place the dot at a desired location and to manipulate the spin. The top gate changes the out of plane electric field $F$ as well as the electron accumulation. The R gate acts as reservoir supplying electrons to the quantum dot. The potential profile is sliced at $z=0$, where the Si/SiO$_2$ interface is located.}
	\label{fig:StepModel}
\end{figure}

Our focus in this work is on the effect of a \textit{single} interface step on the Rabi frequency of an electrically driven spin qubit, such as that depicted in Figure \ref{fig:StepModel}. Due to the large interface electric field, the vertical step creates a sizable potential offset (Fig. \ref{fig:StepModel}b). The electron wave-function moving under the action of an in-plane electric field is initially trapped at the step, but once it acquires enough energy it surmounts the step. \cite{PhysRevB.94.035438} As it does so there is a strong mixing of the orbital and valley degrees of freedom involving all the excited states, and the EDSR frequency goes through a sharp peak as a function of the separation between the step and the center of the dot potential well. This enhancement can be used for fast electrical spin rotations and entanglement even though spin-orbit at silicon interfaces is intrinsically weak, while the sharpness of the peak enables one to suppress spin relaxation by detuning away from it quickly.

This paper is organized as follows. In Sec.~\ref{EDSR} we present the central results of this work and present a method to enhance the EDSR frequency by means of an interface step. The physical implications of the results are discussed in Sec.~\ref{Disc} and their practical applications for device engineering are addressed in Sec.~\ref{appl}. In Sec.~\ref{decoh} we discuss briefly decoherence due to the interplay of roughness and noise. We end with a summary and conclusions.

\section{EDSR near an interface step}
\label{EDSR}

The total Hamiltonian describing the quantum dot-step system~\cite{PhysRevB.84.155320} is $H = H_0 + H_{SOC} + V_c + V_z$. Here $H_0$ is the unperturbed bulk Si Hamiltonian, while $H_{SOC}$ is the spin-orbit Hamiltonian discussed in detail below. The quantum dot is defined by the in-plane confinement potential 
\begin{equation}
	V_c=\frac{\hbar^2}{2m^*(\boldsymbol{r}) a^4}[(x-x_0)^2+y^2]
\end{equation}
 centered at $(x_0,0,0)$, with a radius a = 10 nm and an orbital splitting of 3.8 meV. The effective mass $m^*$ has a longitudinal component $m_z$ ranging from $0.4m_0$ on the SiO$_2$ side to $0.98m_0$ on the Si side, and a transverse component $0.2 m_0$, where $m_0$ is the electron rest mass. The out-of-plane confinement $V_z$ for a flat interface is $V_z=U_0\theta(z)+eFz$, where the hetero-junction barrier potential $U_0\approx3$ eV for Si/SiO$_2$ (150 meV for Si/SiGe). In the presence of a step the interface potential is written as $V_z(x,z)=U_0[\theta(z)\theta(-x)+\theta(z+d)\theta(x)]+eFz$. The step height $d=5.43\AA$ is set to one lattice constant and its location is fixed at $x = 0$. 

Spin-orbit coupling in (001) heterostructures is described generally by the matrices ~\cite{PhysRevB.73.235334,PhysRevB.72.045215}
\begin{equation}
	h_R =\sigma_x k_y - \sigma_yk_x \quad,\quad h_D =\sigma_xk_x-\sigma_yk_y
\end{equation}
where $\sigma_x,\sigma_y$ are spin Pauli matrices, $k_x = -i \frac{\partial}{\partial x}$ and $k_y = -i \frac{\partial}{\partial y}$. The matrix $h_R$ stems from the inversion asymmetry of the confining potential whereas $h_D$ arises from the surface termination. We introduce pseudospin Pauli matrices $\tau_x, \tau_y$ acting in the valley subspace. The total spin-orbit Hamiltonian
\begin{equation}
H_{SOC}=(\alpha \openone+\gamma \tau_y)\otimes h_R+(\beta \openone+\zeta \tau_y)\otimes h_D,
\label{Equation:Interface Potential}
\end{equation}
where $\alpha=5.5\times10^{-14}$ eV cm and $\beta=8\times10^{-14}$ eV cm respectively, and the inter-valley terms $\gamma = 14.3\times10^{-14}$ eV cm and $\zeta = 20.8\times10^{-14}$ eV cm \cite{PhysRevB.92.201401}. 

In the effective mass approximation the electron wave functions~\cite{PhysRev.98.915} $| D_{ns\xi}(x,z) \rangle = \Phi_{n}(x,z)u_\xi{(\textbf{r})}e^{i k_\xi z}\chi(s)$, where $\Phi_{n}(x,z)$ represent the $n$-th level envelope functions and $u_\xi(\textbf{r})$ the lattice periodic Bloch functions corresponding to the valleys centered at $k_\xi={\pm k_0}$~\cite{PhysRevB.80.081305}. The dynamics in the $\hat{\bm y}$-direction are trivial and are neglected henceforth. In the presence of a step the motion in the $\hat{\bm x}$- and $\hat{\bm z}$-directions is no longer separable. The envelope wave function $\Phi_n$ is obtained by solving the effective mass Schr\"odinger equation with the Hamiltonian $H_{EMA} = \hat{p}^2/[2m^*(\boldsymbol{r})] + V_c + V_z(x,z)$ \cite{FRENSLEY1992347} using the Lanczos algorithm on a $160\times275$ finite-element grid. The grid size along the $\hat{\bm z}$-direction is 0.26$\AA$, which captures the effect of atomistic scale interface steps. $\chi(s)$ denotes the spin wave function where $s \in \{\uparrow, \downarrow\}$. The diagonalization results in a relative precision in orbital energy of $1.1\times10^{-3}meV$, and a relative error in the valley splitting of $1.4\times10^{-4}meV$.

For the ground orbital the valley-orbit coupling is 
\begin{equation}
	\Delta_v \displaystyle = \left<D_{0s,\xi}|V_z|D_{0s,\xi'}\right> = \left|\Delta_v\right|e^{-i\phi_v}
\end{equation}
where $\phi_v$ is the mixing phase of the two bare valley states, which in the absence of the step is the same for all orbitals. In the absence of the step EDSR can be captured by a simple perturbative treatment. We restrict our attention to the $8 \times 8$ subspace comprising the ground and first excited orbital states, namely $\{D_{0\uparrow,k},D_{0\downarrow,k},D_{0\uparrow,-k},D_{0\downarrow,-k},D_{1\uparrow,k},D_{1\uparrow,k},D_{1\downarrow,-k},D_{1\downarrow,-k}\}$. The corresponding effective Hamiltonian can be represented as 
\begin{equation}
H_{eff}=\begin{pmatrix}
H_{00} & H_{01}\\
H_{10} & H_{11}
\end{pmatrix}
\end{equation}
The block $H_{00} = E_Z \openone\otimes\sigma_{z} + \Delta_v\tau_{x}\otimes\openone$, with $E_Z$ the Zeeman energy and the ground state orbital energy set to zero, while $H_{11} = \hbar \omega + E_Z  \openone\otimes\sigma_{z}+\Delta_{v,1}\tau_x\otimes\openone$ with $\hbar\omega$ the orbital confinement energy and $\Delta_{v,1} \displaystyle = \left<D_{1s,\xi}|V_z|D_{1s,\xi'}\right>$.
The off-diagonal blocks $H_{01}=H_{10}^\dagger$ represent the matrix elements of the electric dipole interaction and spin-orbit coupling $eE_{ac}x \openone\otimes\openone + H_{SOC}$, with $\beta = \zeta = 0$ for simplicity.

We perform a Schrieffer-Wolff transformation \cite{citeulike:735057,PhysRev.149.491} to project out the $H_{01}$ and $H_{10}$ blocks. Then we diagonalize the resulting matrix to obtain the valley eigenstates $| D_{ns,\pm} \rangle = \frac{1}{\sqrt{2}}(| D_{ns,k_0} \rangle \pm e^{-i\phi_v} | D_{ns,-k_0} \rangle)$, finding a ground state EDSR Rabi frequency ~\cite{Pioro-Ladriere2008,2016arXiv160302829R,PhysRevLett.109.206602}
\begin{equation}
\label{Equation:fRabi_flat}
\begin{array}{l}
f = \frac{g \mu_B eBE_{ac} \kappa \left<x\right>_{01}}{2 \pi \hbar^3 \omega ^2 } \, (\alpha - \gamma \sin \phi _v),
\end{array}
\end{equation}
where $\left<x\right>_{01}$ and $\kappa$ are the matrix elements between the orbital ground and first excited states of the electric dipole and momentum operators respectively.
EDSR arises from two-step virtual processes e.g. $ |D_{0\uparrow, z} \rangle \rightarrow |D_{1\downarrow, -z}\rangle \rightarrow |D_{0\downarrow, -z}\rangle$ and requires spin-orbit coupling, a change in the orbital state, the ac electric field, and time-reversal breaking by the magnetic field. Although the process involves bare valley state mixing, the initial and final valley eigenstates are the same, and the spin states remain in the $|D_{0s,-}>$ subspace.

The resulting effective Hamiltonian for the ground state subspace $\{ D_{0\uparrow,-},D_{0\downarrow,-} \}$ has the form $H_{eff}=\frac{1}{2}\varepsilon_z\sigma_z+\frac{1}{2}\varepsilon_x(t)\sigma_x$, which coincides with the form of the electron spin resonance (ESR) Hamiltonian. Upon application of a vector microwave source, a qubit can be operated around an arbitrary axis on the Bloch sphere via in-phase (X) or in-quadrature (Y) pulses with the reference clock.
\begin{figure}[tbp]
	\includegraphics[width=0.49\linewidth]{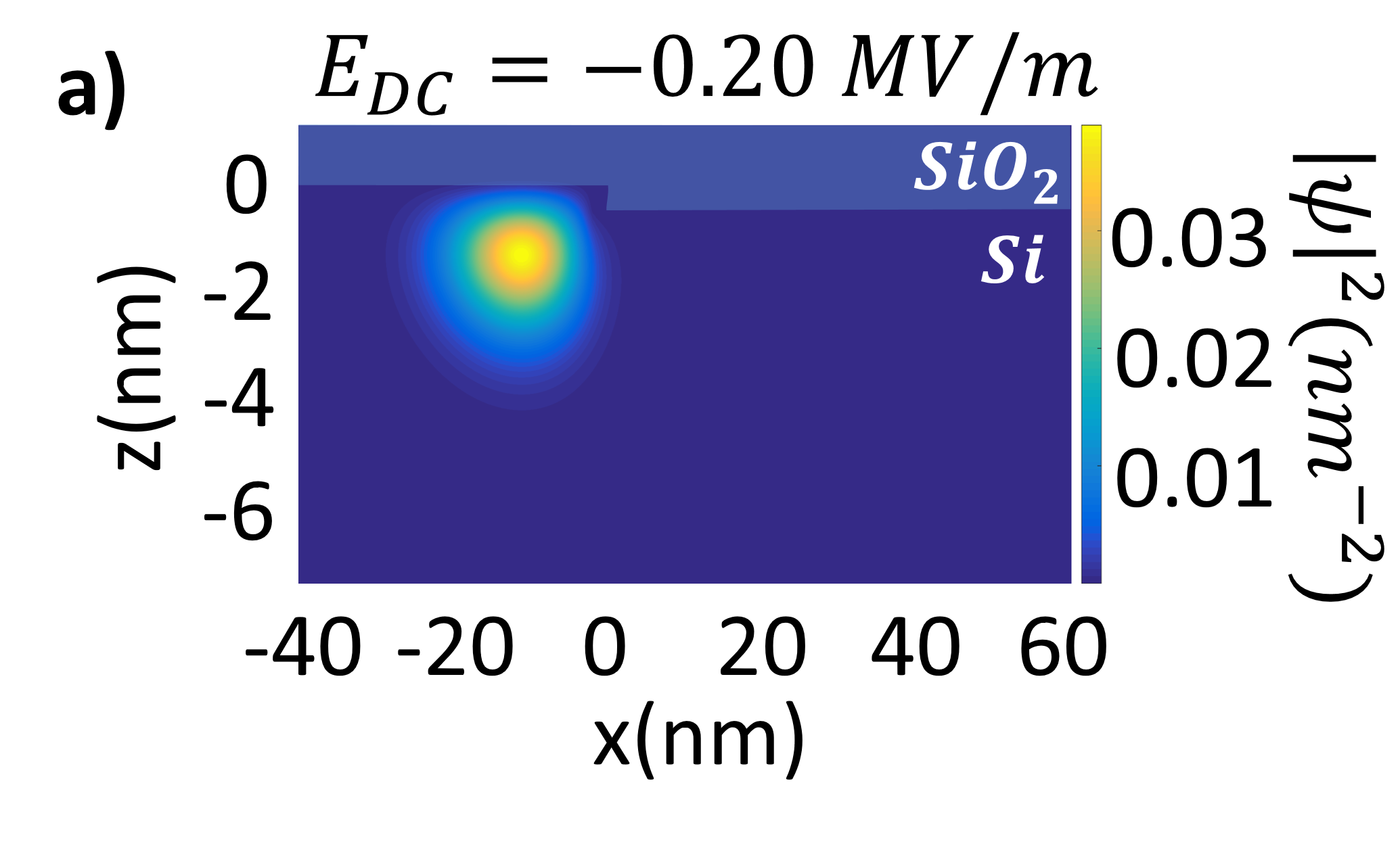}
	\includegraphics[width=0.49\linewidth]{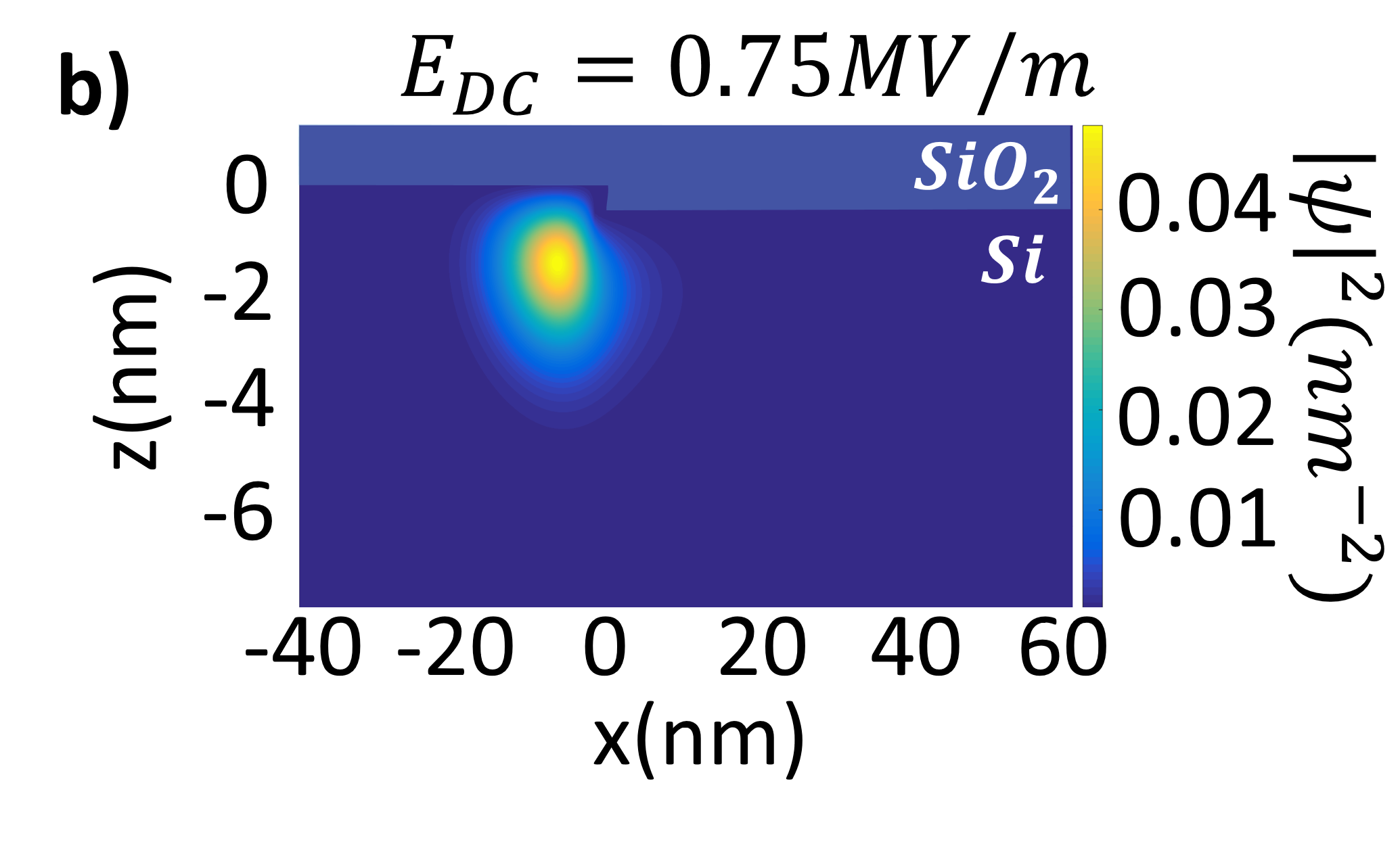}
	\includegraphics[width=0.49\linewidth]{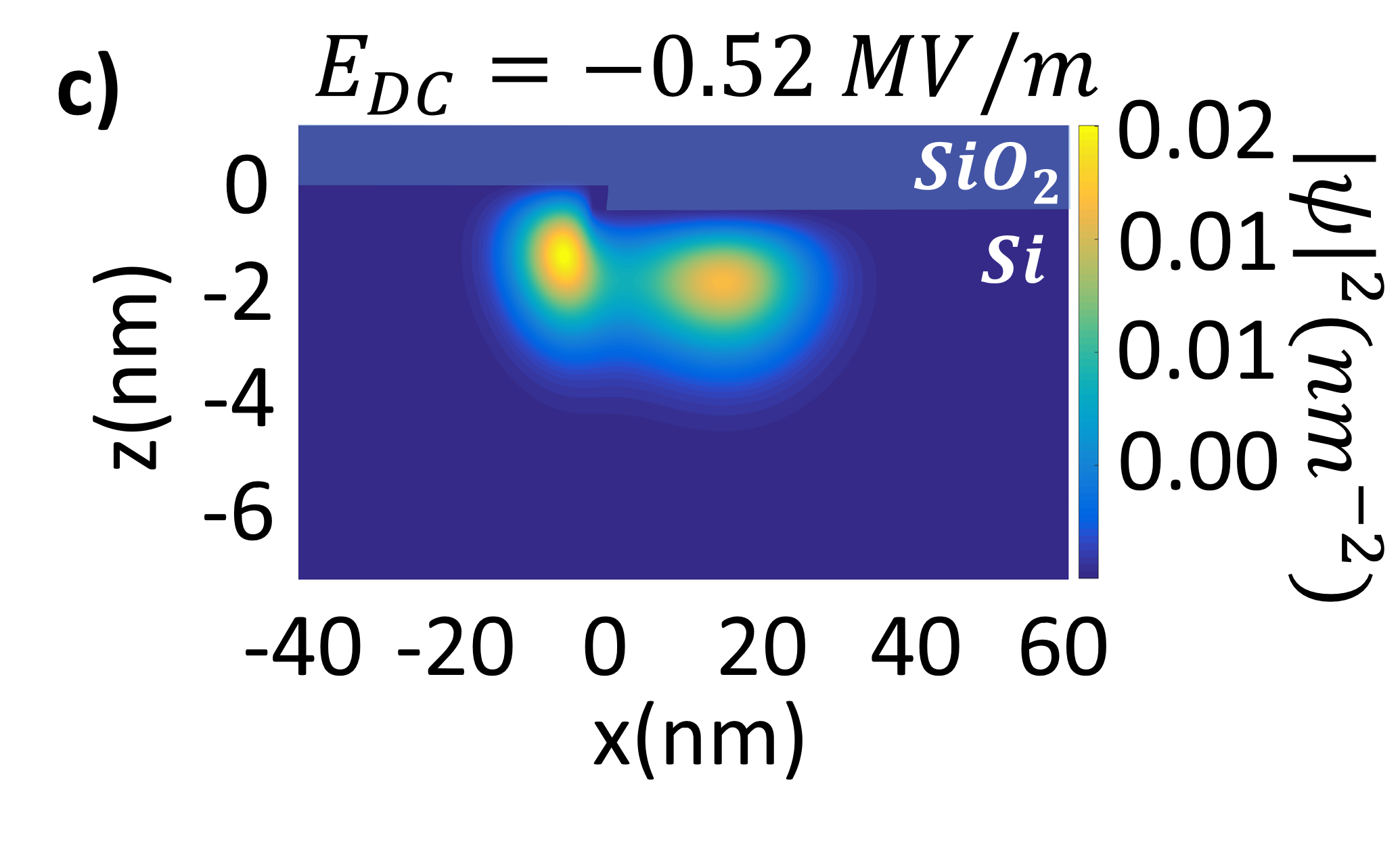}
	\includegraphics[width=0.49\linewidth]{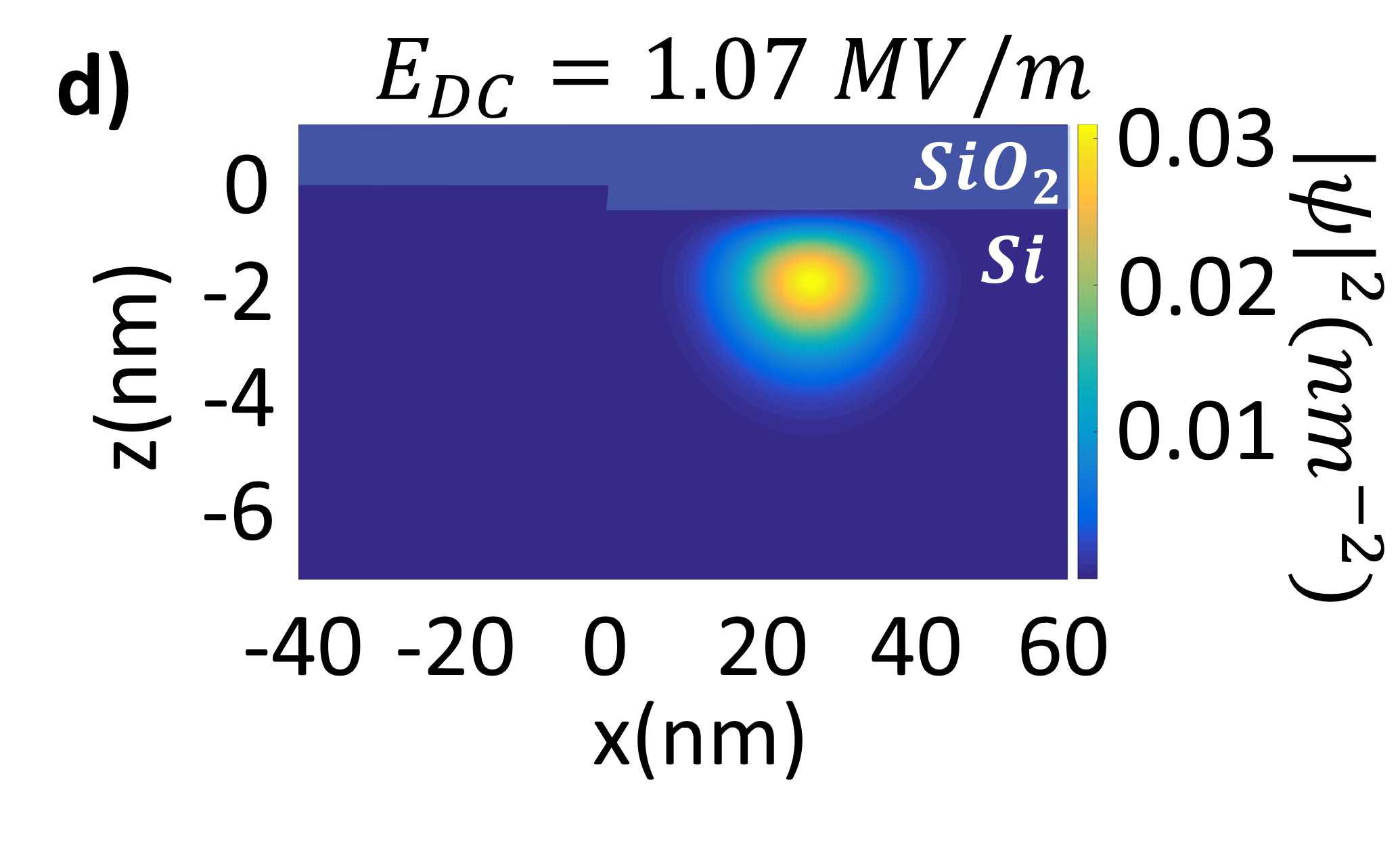}
	\caption{Evolution of the wave function as it is driven over an interface step. The in-plane electric field is used to drag the wave function over a $5.43\AA$ step. During this process (a) The wave function will initially be compressed at the step edge; (b) At higher fields, the electron density starts leaking to the other side of the step; (c) As the wave function pushes against the step, the valley composition become more sensitive to the quantum dot position: in the presence of the step, the in-plane electric field can be used to control both $eE_{ac}<x>_{01}$ and $\Delta_v$, resulting in a significant enhancement of the EDSR frequency. (d) The wave function overcomes the barrier and surmounts the step.}
	\label{fig:Wavefunction}
\end{figure}

\begin{figure}[tbp]
	\includegraphics[width=\linewidth]{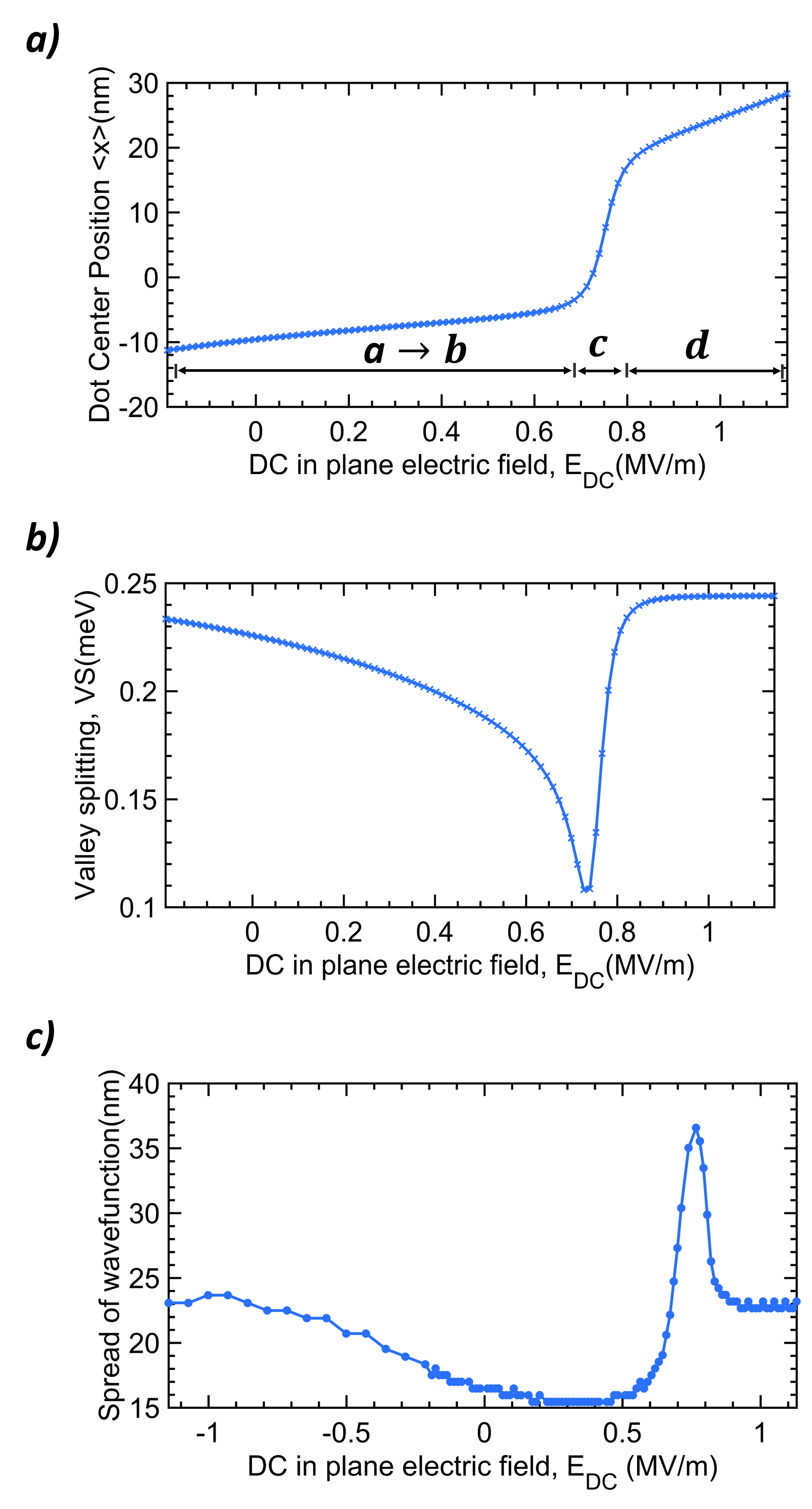}
	\caption{(a) A small DC in-plane electric field ($\approx1MV/m$) due to the side gates shifts the dot ground state mean position $<x>$ by approximately 40 nm. (b) The valley splitting is also sensitive to the location of the center of the dot potential near a $5.43\AA$ step. (c) The spread of the wave function, defined as the extent of the central 90\% of its weight.}
	\label{fig:VS}
\end{figure}
\begin{figure}[tbp]
 	\includegraphics[width=\linewidth]{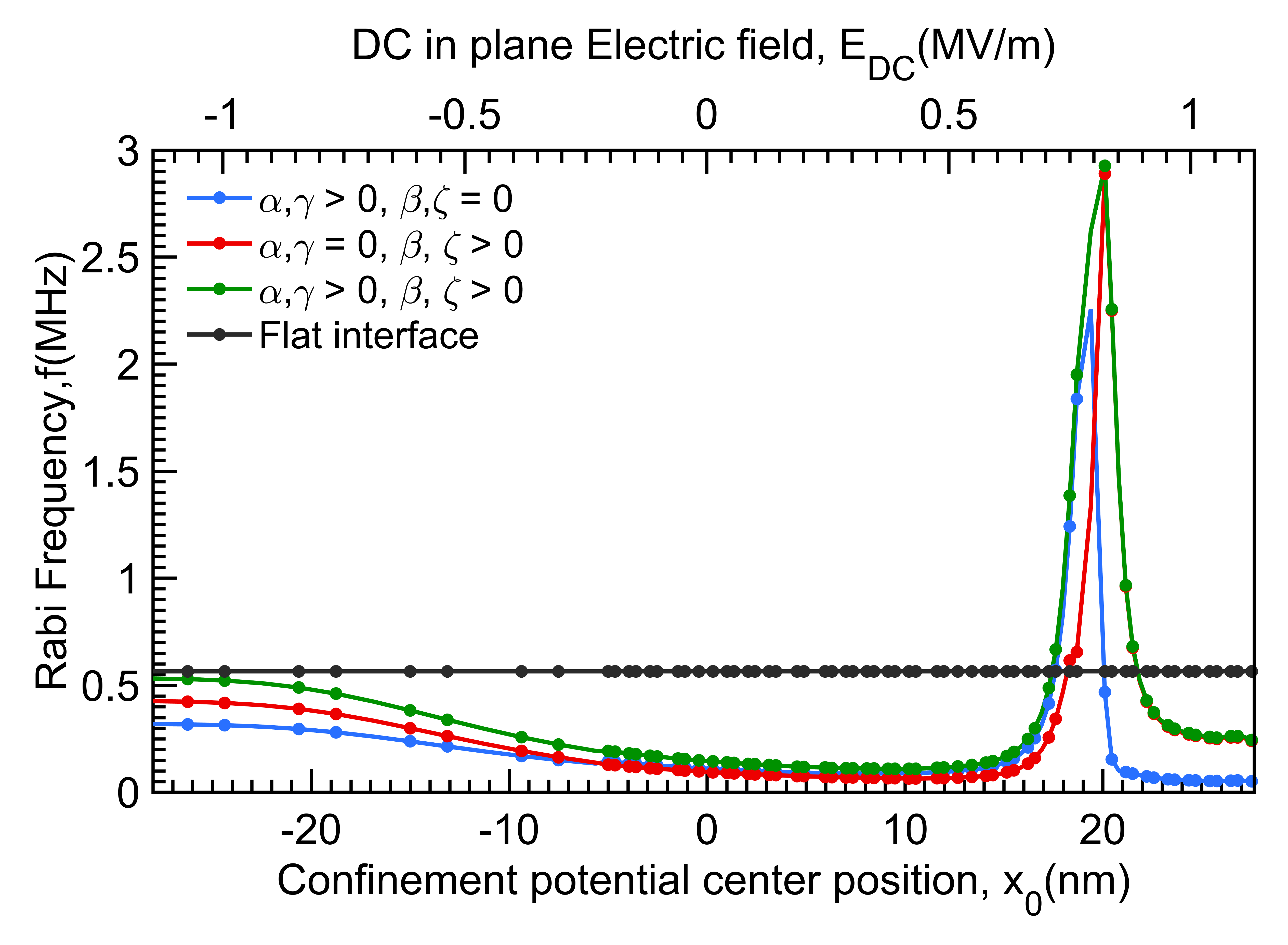}
 	\caption{EDSR Rabi frequency $f_{Rabi,v_-}$ as a function of the quantum dot potential center position $x_0$ in a global magnetic field B=1T along $[001]$ with a step height of $5.43\AA$. The enhancement appears at $x_0=20nm$, on the right side of the step.}
 	\label{fig:5_43SOCBz}
 \end{figure}

In the presence of the step there is a large enhancement of the Rabi frequency because contributions similar to Eq. \ref{Equation:fRabi_flat} arise from the stronger coupling between the ground state (n=0) and the excited states (n=1).  This is indicated by the fact that the wave function extends over a larger area, as higher orbital occupies a wider domain [Fig.\ref{fig:VS}(c)]. Equally importantly, electrical control of the VOC is enabled by the step. Intervalley and intravalley spin-orbit interaction terms couple one valley eigenstate corresponding to the orbital ground state with the opposite valley eigenstate corresponding to the first orbital excited state. The electric field has an additional impact on spin dynamics, leading to a strong enhancement of the Rabi frequency. We take this into account through the spin- and valley-orbit coupling matrix elements ($\left<D_{n,\xi}|H_{SOC}|D_{n',\xi'}\right>$ and $\left<D_{n,\xi}|V_z|D_{n,\xi'}\right>$ respectively) between all pairs of states. Since we work with the exact solution of $H_{EMA}$, the electric dipole term couples the ground state to all excited states. Our numerical results show that an effective Hamiltonian analogous to $H_{eff}$ is sufficient to describe EDSR both quantitatively and qualitatively, the difference being that the individual blocks can no longer be written out in closed form. We determine the Rabi frequency as well as the wave function and its time evolution as a function of the in-plane electric field $E_{ac}$. Only terms linear in $E_{ac}$ are retained. The central result of this paper is displayed in Fig.\ \ref{fig:5_43SOCBz}, which shows the Rabi frequency as a function of the separation between the quantum dot potential center and the step. This is closely related to the evolution of the wave function described in Fig.\ \ref{fig:Wavefunction} and Fig. \ref{fig:VS}(a).

\section{Discussion}
\label{Disc} 

On the far left of Fig.\ \ref{fig:5_43SOCBz} we recover the Rabi frequencies for a flat interface cf. Eq.\ \ref{Equation:fRabi_flat}. These differ slightly on the two sides of the step because, in the presence of the strong interface electric field, the step creates a potential barrier $eFd \approx  15.2 meV$. The valley-orbit coupling magnitude and phase are slightly different on the left and right sides of the step. The potential barrier can be used to understand the sharp, resonance-like shape of Fig.\ \ref{fig:5_43SOCBz}. The spin- and valley-orbit couplings between all the states are maximized at the location of the step. As the wave function approaches the step it initially does not have sufficient energy to overcome it and is pushed against it by the in-plane electric field [Fig.\ref{fig:Wavefunction}(a,b)]. The EDSR Rabi frequency gradually decreases since the additional confinement due to the step limits the movement of the quantum dot ($\left<x\right>_{01}$ decreases as the wave function nears the step). Once the confinement becomes strong enough that the energy of the electron matches that of the step potential barrier, the wave function passes over the step and continues smoothly onto the other side [Fig.\ref{fig:Wavefunction}(c,d)]. As the wave function quickly overcomes the step, the EDSR Rabi frequency has a sharp maximum as a function of position. 

The key to the EDSR enhancement is provided by the intervalley spin-orbit coupling terms governed by the structure-specific parameters $\gamma$ and $\zeta$. The step strongly enhances intervalley dynamics by enabling the electron to tunnel between the ground valley eigenstate and the opposite valley eigenstates corresponding to all excited orbital levels. Thanks to the intervalley spin-orbit coupling, which flips the electron spin, the inter-orbital intervalley tunneling enabled by the step has a strong impact on spin dynamics, and the strong coupling to all the orbital excited states results in a much faster spin rotation than in the absence of the step. This is reflected in the decrease in the valley splitting seen in Fig. \ref{fig:VS}(b). The enhancement of the Rabi frequency is due to the combination of the wave function sensitivity to the in plane electric field as well as to the drop in the valley splitting. 

The enhancement at a step is not present in materials that do not possess a valley degree of freedom, such as III-V semiconductors: we have checked this explicitly. In Si, conversely, the effect is particularly strong since the lowest lying valley states are perpendicular to the interface. Using $E_{ac}$ = 2kV/m and $B$ = 1T we obtain a maximal EDSR gate time $t_{\pi}$ of 170 ns in Si/SiO$_2$, which is approximately five times faster compared to $t_\pi$ = 880 ns for a flat Si/SiO$_2$ interface. In Si/SiGe, a single atomic layer step leads to a peak gate time $t_{\pi}$ of 225 ns, 3 times as fast as for a flat interface. 

Fast qubit operation protects the qubit from unwanted excitations. Given that the spin flip time is $\approx$ 200ns and the orbital state splitting is 3.81meV, our perturbation theory is strongly adiabatic in the orbital motion~\cite{PhysRevLett.109.206602}. Hence, even with the small intrinsic spin-orbit coupling of Si, a spin qubit could be efficiently driven purely by electrical means. A local oscillating electric field allows individual qubit control. Likewise, electrical spin coupling to a superconducting resonator will be enhanced, enabling fast electrical spin entanglement of distinct spin qubits. We note that two-qubit entanglement can also be accomplished via exchange ~\cite{PhysRevA.57.120}. The intervalley spin-orbit terms $\gamma$ and $\zeta$ independently result in an enhancement of the EDSR strength. An additional relative phase may exist between these terms, which is structure-dependent, and slight variations are expected in the EDSR times for individual samples. Yet the effect will be qualitatively the same across all structures and a strong enhancement in EDSR due to the step will occur.

\section{Device application}
\label{appl} 

A typical a.c. electric voltage of approximately 1.5mV was applied to devices in previous EDSR experimental realizations, \cite{Nowack1430,KawakamiE.2014} which we consider to be a realistic voltage representing the current state of the art. By means of Technology Computer Aided Design (TCAD)\cite{TCAD} simulations, we find that in MOS architectures with a silicon oxide thickness of 5nm, a 1.5mV side gate voltage can produce an in plane electric of as much as 2kV/m acting on the quantum dot. This realizes the minimum EDSR gate time $t_{\pi}$ of 170 ns reported above.

\begin{figure}
 	\includegraphics[width=0.95\linewidth]{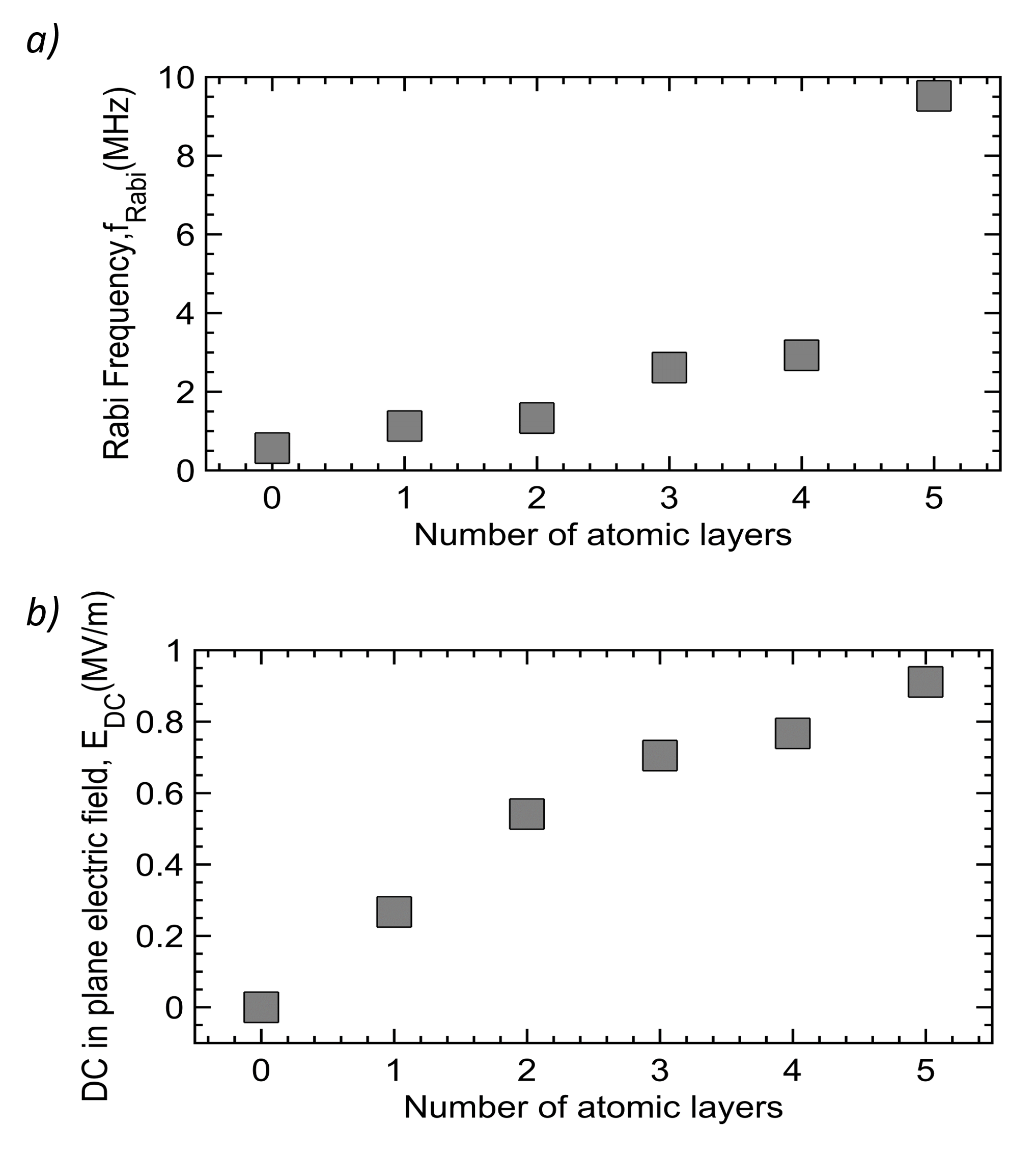}
 	\caption{(a)The enhancement of the EDSR Rabi frequency increases with step height (1 atomic layer=$1.36\AA$).(b) Higher atomic steps require a stronger in-plane DC electric field in order to overcome the potential barrier introduced by the step.}
 	\label{fig:StepHeight}
\end{figure}

The results presented here are crucial for any implementation of EDSR in silicon. Recently, second harmonics have been observed in a Si/SiGe spin qubit operated using EDSR enabled by a nanomagnet \cite{PhysRevLett.115.106802}. A possible explanation of these higher orders can be the presence of disorder, as these cause a strong non-linear dependence of the wave function position and spin-orbit terms on the applied electric field. State-of-the-art technology can reduce Si/SiO$_2$ surface roughness to as low as 0.7 $\AA$ ~\cite{OPL:8064940, OPL:8019159, hiller2010low, PhysRevB.32.8171, anderson1993determination, kim2013charateristics} meaning that only one such step may be present within a single dot device. We also anticipate the possibility of the intentional design of quantum dots incorporating step edges. These could be constructed using standard top-down fabrication techniques, such as reactive ion etching, or possibly STM-based approaches. Such an intentional step would dominate any effects due to interface roughness, as we see in Fig.\ref{fig:StepHeight}(a), which shows that larger steps lead to a stronger enhancement of the Rabi frequency. As the impact of roughness on valley physics averages out~\cite{PhysRevB.82.205315}, the effect of the intentional step becomes more prominent. However, as the step increases in size, a stronger DC in plane electric field is required to push the center of the quantum dot to a position where the spin can reach the maximum possible EDSR Rabi frequency. As the step size is increased by the addition of further atomic layers, it becomes increasingly difficult for the electron to overcome the potential due to the step [Fig.\ref{fig:StepHeight}(b)]. For large steps, which prevent the formation of a simple, single quantum dot, the potential landscape eventually becomes rather complicated. 

\section{Decoherence}
\label{decoh}

The step may increase the coupling to phonons and charge noise, causing spin and valley relaxation~\cite{Yang2013, 2016arXiv160302829R}. Since the qubit is in the lowest valley eigenstate, valley relaxation will only be important around hotspots. Yet the intervalley spin-orbit coupling could enhance decoherence mechanisms already active in the absence of valley-orbit coupling \cite{:/content/aip/journal/apl/105/19/10.1063/1.4901162,PhysRevB.76.245322,PhysRevB.76.195204,PhysRevB.91.165432, Nowack1430,PhysRevB.71.205324,PhysRevLett.93.016601,PhysRevB.83.165322,PhysRevB.71.075315}. Moreover, interface roughness is unavoidable in heterostructures, giving rise to fluctuations in the $z$-position of the interface that couple different valley eigenstates~\cite{PhysRevB.82.205315}. Noise and phonons driving the quantum dot over the fast-varying roughness profile may enable intervalley tunnelling. Together with the intervalley spin-orbit coupling this may lead to additional spin relaxation and dephasing. 

Nevertheless, the sharpness of the resonance in Fig.\ \ref{fig:5_43SOCBz} means that experimentally the quantum dot position only needs to be tuned 5 nm away from the step once the spin rotation is accomplished for the spin relaxation and dephasing times to return to their normal values for a flat interface. Charge noise and phonons will only be noticeable during qubit operation. To preserve fidelity experiment should ensure the qubit is at the maximum in Fig.\ \ref{fig:5_43SOCBz}, where the sensitivity to jitter is eliminated. Moreover, roughness will reduce the magnitude of the valley-orbit coupling (i.e. the valley splitting), yet as long as the valley splitting can be resolved experimentally the spin dynamics described in this work should be observable. 
 
\section{Conclusions and outlook} 

We have demonstrated that a single step at a silicon heterointerface strongly enhances EDSR in a single-spin qubit. The effect is driven by intervalley spin-orbit coupling terms specific to silicon, and by the intervalley tunnelling enabled by the step. The Rabi frequency has a sharp maximum as a function of the qubit position, such that the qubit can be tuned away from the step to reduce spin relaxation and dephasing. A high gate fidelity can be maintained by positioning the qubit at the location that yields the maximum EDSR frequency. Our findings pave the way for the experimental realization of EDSR in silicon without a nanomagnet, despite spin-orbit coupling being inherently weak. 

\acknowledgements
We thank Andr\'as Palyi, Leonid Golub and Andrea Morello for enlightening discussions. The authors acknowledge support from the Australian Research Council (Grant No. CE110001027), the U.S. Army Research Office (Grant No. W911NF-13-1-0024), and the NSW Node of the Australian National Fabrication Facility.  M.V. acknowledges support from the Netherlands Organization for Scientific Research (NWO) through a Rubicon Grant.
\bibliography{mybib}
\appendix
\section{Form of Spin-orbit coupling}

The spin-orbit coupling in the basis $\{D_{0\uparrow_,k_0},D_{0\downarrow_,k_0},D_{0\uparrow_,-k_0},D_{0\downarrow_,-k_0}\}$ has the form
\begin{equation}
\label{Eq:SOCMatirx}
H_{SOC}=\left(
\begin{array}{cc} 

H_{D_{intra}}+H_{R_{intra}} &
H_{D_{inter}}+H_{R_{inter}}\\[3ex]
H_{D_{inter}}^* + H_{R_{inter}}^  * & H_{D_{intra}}+H_{R_{intra}}
\\
\end{array}
\right)
\end{equation}
In Eq.\ref{Eq:SOCMatirx}, the intra-valley Rashba terms have the form
\begin{equation}
H_{R_{intra}}=\alpha \left(k_y  \sigma _x-k_x  \sigma _y\right)=\left(
\begin{array}{cc}
0 & i \alpha  k_- \\
-i \alpha  k_+ & 0 \\
\end{array}
\right),
\end{equation}
where $k_{\pm}=k_x\pm ik_y$. The inter-valley Rashba terms can be written as
\begin{equation}
H_{R_{\text{inter}}}=-i \gamma \left(k_y  \sigma _x-k_x  \sigma _y\right)=\left(
\begin{array}{cc}
0 & \gamma  k_- \\
-\gamma  k_+ & 0 \\
\end{array}
\right).
\end{equation}
The intra-valley Dresselhaus spin-orbit coupling has the form
\begin{equation}
H_{D_{intra}}=\beta \left(k_x  \sigma _x-k_y 
\sigma _y\right)=\left(
\begin{array}{cc}
0 & \beta  k_+ \\
\beta  k_- & 0 \\
\end{array}
\right),
\end{equation}
with the inter-valley terms
\begin{equation}
H_{D_{inter}}=\zeta \left(k_x\sigma _x-k_y\sigma _y\right)=\left(
\begin{array}{cc}
0 & \zeta  k_+ \\
\zeta  k_- & 0 \\
\end{array}
\right)
\end{equation}
The magnetic field is applied along [001], corresponding to a Zeeman interaction
\begin{equation}
H_z=\frac{g\mu_B B}{2}\left(
\begin{array}{cc}
1 & 0 \\
0 & -1 \\
\end{array}
\right).
\end{equation}

\section{Coupling matrix elements}

The envelope wave function for the excited state is $\Phi_1(x,z)=\frac{1}{a^2\sqrt{\pi}}(x-X_D)e^{-\frac{(x-X_D)^2}{2 a^2}}\psi(s)$, giving rise to the following matrix elements
\begin{equation}
\begin{array}{ll}
<k_x>_{01} &=\frac{-i}{a^3\pi}\int\limits_{-\infty}^{\infty}dx e^{-\frac{(x-X_D)^2}{2 a^2}}(\frac{\partial}{\partial x})(x-X_D)e^{-\frac{(x-X_D)^2}{2 a^2}}\\[2ex]
&= -\frac{i}{\sqrt{2}a}=-i \kappa\\[3ex]
<k_y>_{01} &=0\\
\end{array}
\end{equation}
Similarly the matrix element for $<x>_{01}$ is
\begin{equation}
\begin{array}{ll}
<x>_{01} & =\xi=\frac{1}{a^3\pi}\int\limits_{-\infty}^{\infty}dx e^{-\frac{(x-X_D)^2}{2 a^2}}(x-X_D)e^{-\frac{(x-X_D)^2}{2 a^2}}\\[3ex]
&= \frac{a}{\sqrt{2}}\\
\end{array}
\end{equation}
Thus the total Hamiltonian in the basis $\{D_{0\uparrow_,k_0},D_{0\downarrow_,k_0},D_{0\uparrow_,-k_0},D_{0\downarrow_,-k_0},D_{1\uparrow_,k_0},D_{1\downarrow_,k_0},D_{1\uparrow_,-k_0},D_{1\downarrow_,-k_0}\}$ with Rashba only spin-orbit coupling becomes
\begin{equation} 
\begin{array}{l}
H=\\[3ex]
\begin{psmallmatrix}
\frac{E_z}{2} & 0 & \Delta_v & 0 & eE\xi & \alpha  \kappa  & 0 & -i \gamma  \kappa  \\
0 & -\frac{E_z}{2} & 0 & \Delta_v & -\alpha  \kappa  & eE\xi & i \gamma  \kappa  & 0 \\
\Delta_v^* & 0 & \frac{E_z}{2} & 0 & 0 & i \gamma  \kappa  & eE\xi & \alpha  \kappa  \\
0 & \Delta_v^* & 0 & -\frac{E_z}{2} & -i \gamma  \kappa  &0 & -\alpha  \kappa  & eE\xi \\
eE\xi & -\alpha  \kappa  & 0 & i \gamma  \kappa  & \frac{E_z}{2}+\hbar \omega    & 0 & \Delta_v & 0 \\
\alpha  \kappa  & eE\xi & -i \gamma  \kappa  & 0 & 0 &  -\frac{E_z}{2}+ \hbar\omega  & 0 & \Delta_v \\
0 & -i \gamma  \kappa  & eE\xi & -\alpha  \kappa  & \Delta_v^* & 0 & \frac{E_z}{2}+ \hbar \omega   & 0 \\
i \gamma  \kappa  & 0 & \alpha  \kappa  & eE\xi & 0 & \Delta_v^* & 0 & -\frac{E_z}{2}+\hbar \omega   \\[3ex]
\end{psmallmatrix}
\end{array}
\end{equation}

We apply the Schrieffer-Wolff transformation to the off-diagonal elements of $H$, ignoring all the higher order spin-orbit terms $\alpha ^2 \kappa ^2+\gamma ^2 \kappa^2$ and $\alpha  \gamma \kappa ^{2}$. We obtain an energy offset for ground state subspace $\{D_{0\uparrow_,k_0},D_{0\downarrow_,k_0},D_{0\uparrow_,-k_0},D_{0\downarrow_,-k_0}\}$ 
\begin{equation}
\begin{array}{l}
H^{(2)}=\\[2ex]
\begin{psmallmatrix}
0 & -\frac{ \alpha g \mu _B e B   E \kappa \xi }{\hbar ^2 \omega ^2 } & 0 & \frac{i \gamma  g \mu _B e  B E   \kappa \xi }{\hbar ^2 \omega ^2 } \\
-\frac{ \alpha g \mu _B e B   E \kappa \xi }{\hbar ^2 \omega ^2 } & 0 & \frac{i \gamma  g \mu _B e  B E   \kappa \xi }{\hbar ^2 \omega ^2 } & 0 \\
0 & -\frac{i \gamma  g \mu _B e  B E   \kappa \xi }{\hbar ^2 \omega ^2 } & 0 & -\frac{\alpha g \mu _B e B   E \kappa \xi }{\hbar ^2 \omega ^2 } \\
-\frac{i \gamma  g \mu _B e  B E   \kappa \xi }{\hbar ^2 \omega ^2 } & 0 & -\frac{\alpha g \mu _B e B   E \kappa \xi }{\hbar ^2 \omega ^2 } & 0 \\[3ex]
\end{psmallmatrix}
\end{array}
\end{equation}
As an example of the calculation, the matrix element (1,2) is
	\begin{equation}
	\begin{array}{ll}
	H^{(2)}_{12} & =\frac{1}{2}\sum_{l} H_{1l}H_{l2}[\frac{1}{E_1-E_l}+\frac{1}{E_2-E_l}]\\[2ex]
%	&=\frac{1}{2}(H_{15}H_{52})[\frac{1}{E_1-E_5}+\frac{1}{E_2-E_5}]+\frac{1}{2}(H_{16}H_{62})[\frac{1}{E_1-E_6}+\frac{1}{E_2-E_6}]\\[2ex]
%	&=\frac{1}{2}[eE\xi (-\alpha  \kappa)][\frac{1}{\frac{g\mu_B B}{2}-(\frac{g\mu_B B}{2}+\hbar \omega)}+\frac{1}{-\frac{g\mu_B B}{2}-(\frac{g\mu_B B}{2}+\hbar \omega)}]+\frac{1}{2}(\alpha  \kappa eE\xi)[\frac{1}{\frac{g\mu_B B}{2}-(-\frac{g\mu_B B}{2}+\hbar \omega)}+\frac{1}{-\frac{g\mu_B B}{2}-(-\frac{g\mu_B B}{2}+\hbar \omega)}]\\[2ex]
%	&=\frac{1}{2}(\alpha eE\xi \kappa)[\frac{1}{\hbar \omega}+\frac{1}{g\mu_B B+\hbar \omega}]+\frac{1}{2}(\alpha eE\xi \kappa)[\frac{1}{g\mu_B B-\hbar \omega}-\frac{1}{\hbar \omega}]\\[2ex]
	&=\frac{1}{2}(\alpha eE\xi \kappa)[\frac{1}{g\mu_B B+\hbar \omega}+\frac{1}{g\mu_B B-\hbar \omega}]\\
	
	\end{array}
	\end{equation}
We consider the valley orbit coupling in the ground state subspace
\begin{equation}
\begin{array}{l}
H=\\[2ex] 
\begin{psmallmatrix}
\frac{E_z}{2} & -\frac{ \alpha g \mu _B e B   E \kappa \xi }{\hbar ^2 \omega ^2 } & \Delta_v & \frac{i \gamma  g \mu _B e  B E   \kappa \xi }{\hbar ^2 \omega ^2 } \\
-\frac{ \alpha g \mu _B e B   E \kappa \xi }{\hbar ^2 \omega ^2 } & -\frac{E_z}{2} & \frac{i \gamma  g \mu _B e  B E   \kappa \xi }{\hbar ^2 \omega ^2 } & \Delta_v \\
\Delta_v^* & -\frac{i \gamma  g \mu _B e  B E   \kappa \xi }{\hbar ^2 \omega ^2 } & \frac{E_z}{2} & -\frac{\alpha g \mu _B e B   E \kappa \xi }{\hbar ^2 \omega ^2 } \\
-\frac{i \gamma  g \mu _B e  B E   \kappa \xi }{\hbar ^2 \omega ^2 } & \Delta_v^* & -\frac{\alpha g \mu _B e B   E \kappa \xi }{\hbar ^2 \omega ^2 } & -\frac{E_z}{2} \\
\end{psmallmatrix}
\end{array}
\end{equation}
We diagonalize this matrix using the rotation
\begin{equation}
R=\frac{1}{\sqrt{2}}\left(
\begin{array}{cccc}
1 & 0 & e^{-i \phi _v} & 0 \\
0 & 1 & 0 & e^{-i \phi _v} \\
1 & 0 & -e^{-i \phi _v} & 0 \\
0 & 1 & 0 & -e^{-i \phi _v} \\
\end{array}
\right)
\end{equation}
yielding
\begin{equation}
R H R^{-1}=\left(\begin{array}{cc}
H_{v_+} & H_{01}\\
H_{10} & H_{v_-}
\end{array}\right)
\end{equation}
where
\begin{equation}H_{v_+}=\left(
\begin{array}{cc}
	|\Delta_v| +\frac{1}{2} E_z & -\frac{g \mu _B  e B E \kappa  \xi \left[\alpha +\gamma \sin \left(\phi _v\right)\right]}{\hbar ^2  \omega ^2}\\[2ex]
	-\frac{g \mu _B  e B E \kappa  \xi \left[\alpha +\gamma \sin \left(\phi_v\right)\right]}{\hbar ^2  \omega ^2} & |\Delta_v| -\frac{1}{2} E_z  \\[2ex]
\end{array}\right)
\end{equation}
and
\begin{equation}
H_{01}=\left(\begin{array}{cc}
0 & -\frac{i B e g \gamma E \kappa \mu _B \xi \cos \left(\phi _v\right)}{\hbar ^2  \omega ^2} \\[2ex]
-\frac{i B e g \gamma E \kappa \mu _B \xi \cos \left(\phi _v\right)}{\hbar ^2  \omega ^2} &0 \\[2ex]
\end{array}\right)
\end{equation}
as well as
\begin{equation}
H_{v_-}=\left(\begin{array}{cc}
-|\Delta_v|+\frac{1}{2} E_z & \frac{g \mu _B  e B E \kappa  \xi \left[\gamma \sin \left(\phi _v\right)-\alpha \right]}{\hbar ^2  \omega ^2} \\[2ex]
\frac{g \mu _B  e B E \kappa  \xi \left[\gamma \sin \left(\phi_v\right)-\alpha \right]}{\hbar ^2  \omega ^2} & -|\Delta_v| -\frac{1}{2} E_z \\[2ex]
\end{array}\right).
\end{equation}
This finally yields the Rabi frequency in the subspace spanned by the spin-split ground valley eigenstate
\begin{equation}
\begin{array}{l}
f = \frac{g \mu_B eBE_{ac} \kappa \left<x\right>_{01}}{2 \pi \hbar^3 \omega ^2 } \, (\alpha - \gamma \sin \phi _v).
\end{array}
\end{equation}
\end{document}